\documentclass{article}

\PassOptionsToPackage{numbers, compress}{natbib}
\usepackage[dblblindworkshop,final]{neurips_2025}
\workshoptitle{The Second Workshop on GenAI for Health: Potential, Trust, and Policy Compliance}
\usepackage[utf8]{inputenc} % allow utf-8 input
\usepackage[T1]{fontenc}    % use 8-bit T1 fonts
\usepackage{hyperref}       % hyperlinks
\usepackage{url}            % simple URL typesetting
\usepackage{booktabs}       % professional-quality tables
\usepackage{amsfonts}       % blackboard math symbols
\usepackage{nicefrac}       % compact symbols for 1/2, etc.
\usepackage{microtype}      % microtypography
\usepackage{xcolor}         % colors
\usepackage{graphicx}
\usepackage{booktabs}

\title{Zero-Shot Large Language Model Agents for Fully Automated Radiotherapy Treatment Planning}

\author{
Dongrong Yang$^{1}$, Xin Wu$^{1}$, Yibo Xie$^{1}$, Xinyi Li$^{2}$, Qiuwen Wu$^{1}$, Jackie Wu$^{1}$, Yang Sheng$^{1}$\thanks{Corresponding author.}\\
$^{1}$Department of Radiation Oncology, Duke University, Durham, NC, USA\\
$^{2}$Department of Radiation Oncology, University of California Irvine, Irvine, CA, USA\\
\texttt{dongrong.yang@duke.edu, yang.sheng@duke.edu}
}

\begin{document}

\maketitle
\begin{abstract}
  Radiation therapy treatment planning is an iterative, expertise-dependent process, and the growing burden of cancer cases has made reliance on manual planning increasingly unsustainable, underscoring the need for automation.In this study, we propose a workflow that leverages an large language model (LLM)-based agent to navigate inverse treatment planning for intensity-modulated radiation therapy (IMRT). The LLM agent was implemented to directly interact with a clinical treatment planning system (TPS) to iteratively extract intermediate plan states and propose new constraint values to guide inverse optimization. The agent’s decision-making process is informed by current observations and previous optimization attempts and evaluations, allowing for dynamic strategy refinement. The planning process was performed in a zero-shot inference setting, where the LLM operated without prior exposure to manually generated treatment plans and was utilized without any fine-tuning or task-specific training. The LLM-generated plans were evaluated on twenty head-and-neck cancer cases against clinical manual plans with key dosimetric endpoints' statistics analyzed and reported. LLM-generated plans achieved comparable organ-at-risk (OAR) sparing relative to clinical plans, while demonstrating improved hot spot control ($D_{\mathrm{max}}$: 106.5\% vs.\ 108.8\%) and superior conformity (conformity index: 1.18 vs.\ 1.39 for boost PTV; 1.82 vs.\ 1.88 for primary PTV). This study demonstrates the feasibility of a zero-shot, LLM-driven workflow for automated IMRT treatment planning in a commercial TPS. The proposed approach provides a generalizable and clinically applicable solution that could reduce planning variability and support broader adoption of AI-based planning strategies.
\end{abstract}

\section{Introduction}

Radiotherapy remains an essential component of modern cancer management, with evidence-based models indicating that approximately 50--70\% of patients should receive at least one course during their disease trajectory, representing millions of new cases globally each year \citep{RN1,RN2}. Global projections forecast a substantial escalation in demand through 2050, further intensifying pressure on already constrained planning resources \citep{RN3}. In parallel, workforce analyses highlight that this strain extends directly to treatment planning personnel: a national survey of the UK dosimetrist workforce reported planning as the predominant professional responsibility, coupled with recruitment challenges and training bottlenecks that threaten service capacity \citep{RN4}. Complementary U.S.\ data document persistent shortages in the medical physics workforce, a core contributor to treatment planning, with implications for plan turnaround times and quality assurance \citep{RN5}. Similar trends have been observed across Europe, where the ESTRO-HERO study identified widespread staffing deficits within radiotherapy departments \citep{RN6}. In the context of IMRT and volumetric modulated arc therapy (VMAT), manual planning remains labor-intensive and prone to substantial inter-planner variability in target coverage and organ-at-risk (OAR) sparing, even within the same institution \citep{RN7,RN8}.

Consequently, automation of treatment planning has been a long-standing priority within the radiation oncology community, aimed at mitigating workload pressures and reducing inter-planner variability. Broadly, current approaches can be categorized into four paradigms: knowledge-based planning (KBP), which leverages historical plan libraries to predict achievable dose--volume histograms (DVHs) to guide optimization \citep{RN9,RN10,RN11,RN12,RN13,RN14}; protocol-based planning, which applies standardized objective sets and optimization priorities to generate consistent plans \citep{RN15,RN16,RN17,RN18}; multi-criteria optimization (MCO), which allows planners to navigate trade-offs between competing objectives \citep{RN19,RN20,RN21,RN22}; and reinforcement learning (RL), in which artificial agents iteratively adjust plan parameters based on dosimetric feedback \citep{RN23,RN24,RN25,RN26,RN27,RN28}. Each approach offers distinct advantages: KBP can embed institutional expertise, protocol-based planning enforces consistency, MCO provides transparent trade-off control, and RL enables adaptive decision-making. However, all suffer from limitations that restrain broad clinical adoption: KBP often require large, high-quality labeled datasets; protocol-based methods lack flexibility for complex or atypical anatomies; MCO demands substantial planner engagement and expertise; and RL approaches can be computationally intensive and require expert-crafted reward functions. 

As a result, despite decades of progress, few auto-planning solutions have achieved universal applicability in routine clinical workflows, underscoring the need for more adaptable and generalizable solutions. Large language models (LLMs) have progressed rapidly in recent years, with successive generations achieving remarkable improvements in reasoning and generalization through increased scale and training data diversity \citep{RN29,RN30}. While early LLMs struggled with tasks outside their training distribution, scaling studies have shown that beyond certain thresholds, these models exhibit abrupt gains in capability, including zero-shot and few-shot learning \citep{RN31,RN32,RN33}. Such capabilities are particularly relevant to radiation therapy treatment planning, a highly specialized domain with complex decision-making requirements and scarce publicly available training data. By leveraging zero-shot and few-shot capabilities, an LLM provided with explicit clinical objectives and integrated into the treatment planning system (TPS) could adapt its reasoning to diverse clinical scenarios without the need for extensive retraining. This learning paradigm offers a distinct advantage over existing approaches, as it does not rely on large expert-labeled datasets or highly engineered domain-specific systems, thereby facilitating broader generalization and transferability once an operational workflow is established.

In this study, we designed and implemented a feasible workflow for leveraging an LLM agent to generate radiation therapy treatment plans entirely in a zero-shot manner, without reliance on prior plans. To capitalize on the LLM’s general reasoning capabilities, the complex treatment planning task was decomposed into domain-agnostic subtasks, guided by clinical objectives and broadly applicable planning principles. The agent autonomously extracted intermediate plan states (e.g., DVHs, objective function losses, and dose–volume objectives), analyzed these states using arithmetic and trend-based reasoning, and iteratively proposed updated optimization objectives. Feasibility was demonstrated in 20  head-and-neck (HN) IMRT cases, with key dosimetric endpoints compared against corresponding clinical plans.

\section{Methods}
\label{sec:methods}

\subsection{Dataset and Planning Environment}

To ensure clinical relevance and broad applicability, the proposed workflow was developed and validated on a widely adopted commercial TPS Eclipse™ (version 15.6, Varian Medical Systems, Palo Alto, CA). The LLM-based agent was designed to iteratively adjust treatment planning parameters within the inverse optimization space to meet clinical objectives and enhance plan quality.
The agent interfaces directly with the Eclipse TPS via the Eclipse Scripting Application Programming Interface (ESAPI), enabling programmatic access to the treatment planning environment. Through ESAPI, the agent can retrieve intermediate planning states (e.g., DVH metrics, objective function values) and modify inverse planning constraints in a manner similar to that of a human planner. This integration ensures that all interactions occur within the native TPS environment, maintaining consistency with clinical workflows and eliminating discrepancies that may arise from surrogate optimization engines or approximated planning platforms.
We validated the feasibility of the proposed method by applying it to IMRT treatment planning for HN cancer, a particularly challenging site due to the close proximity of multiple critical OARs. The substantial anatomical overlap between the planning target volume (PTV) and nearby OARs necessitates complex, non-intuitive trade-off decisions during plan optimization. The agent was tasked with analyzing the planning status and navigating patient-specific trade-offs to optimize plan quality for each case. Twenty HN patients received IMRT treatment in our institution were retrospectively collected with institutional IRB approval.  All patients received the same prescription regimen: 70~Gy to the boost planning target volume (PTV$_{\mathrm{boost}}$ ) and 44~Gy to the primary PTV (PTV$_{\mathrm{primary}}$), delivered in 2~Gy per fraction. The clinical plans were manually generated by certified dosimetrist and the dose distribution was reviewed and approved by the attending radiation oncologist prior to treatment delivery. For automated plan generation, the same target and OAR contours, as well as prescription dose levels, were used to ensure consistency and enable direct comparison with the corresponding clinical plans.

\subsection{LLM-based Agentic Workflow for Automatic Inverse Planning}
\begin{figure}[htbp]
  \centering
  \includegraphics[width=0.9\linewidth]{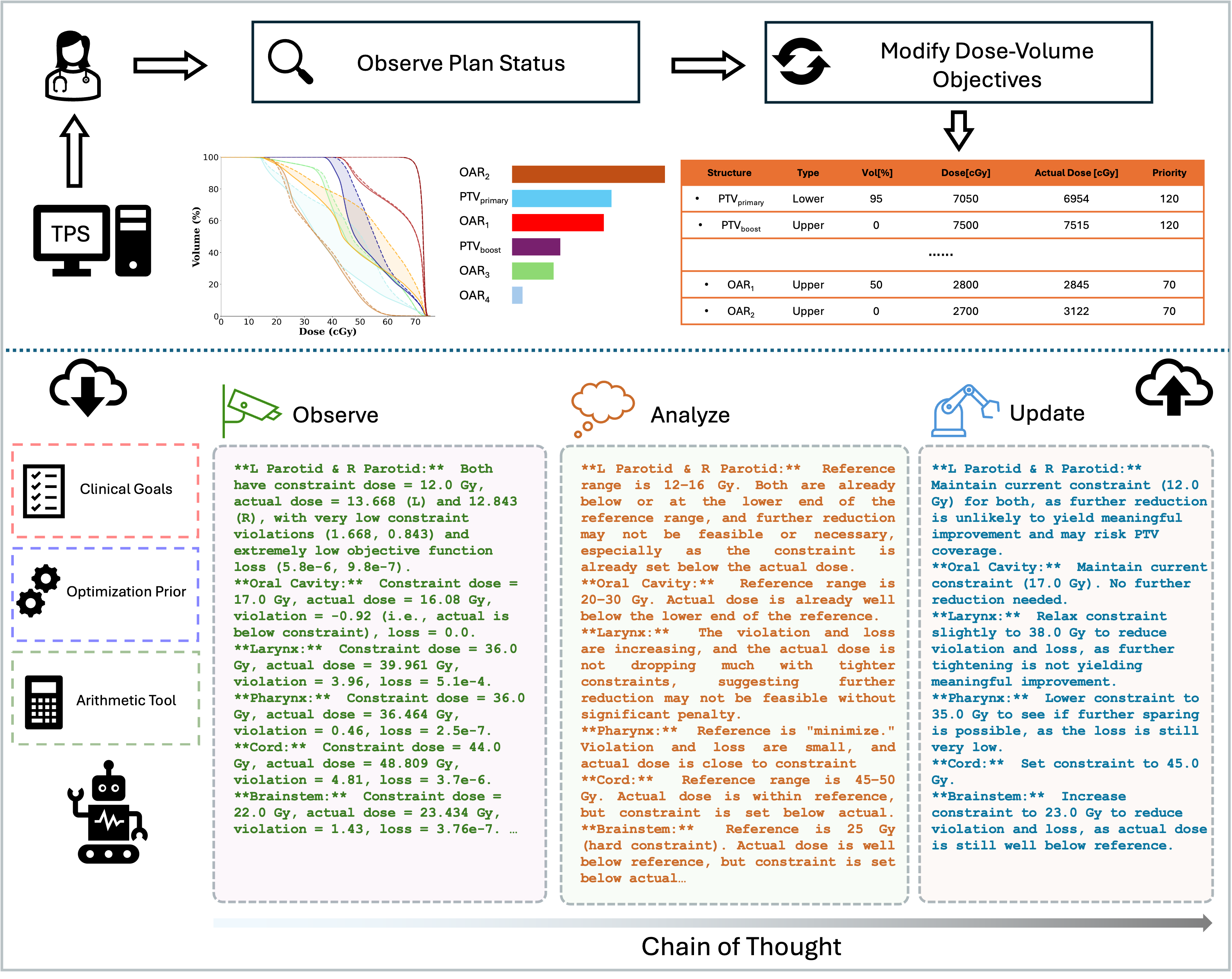}
  \caption{LLM-based agentic workflow for automatic inverse planning. The upper panel illustrates the conventional manual planning workflow, where a human planner iteratively reviews intermediate plan states and adjusts dose–volume constraints. The lower panel depicts the proposed LLM-driven agentic workflow, designed to mimic the manual process. Guided by clinical objectives, prior knowledge of optimization systems, and access to computational tools, the LLM leverages its general reasoning capability to analyze plan status and adapt constraints in a human-like manner. At each iteration, structured chain-of-thought reasoning is applied to enhance decision quality and constraint refinement}
  \label{fig:workflow}
\end{figure}

The central concept of the proposed workflow is to leverage an LLM-based agent to iteratively refine optimization objectives during the inverse treatment planning process, with the goal of producing high-quality treatment plans. As shown in Figure ~\ref{fig:workflow}, the workflow consists of two key components. First, the LLM agent is designed to directly interact with the TPS, allowing it to extract relevant planning information and adjust plan parameters to guide optimization. Second, informed by the current plan status, the agent applies its general reasoning capabilities to propose clinically meaningful modifications that can effectively improve overall plan quality.

To support the LLM's decision-making process, we drew inspiration from standard manual planning workflows. In clinical practice, planners iteratively adjust optimization constraints based on their observations of key dosimetric endpoints extracted from DVH curves, as well as objective function feedback from the optimization engine. The dosimetric endpoints serve as clinically relevant indicators, quantifying how the current dose distribution aligns with prescribed clinical goals. In parallel, the objective function loss is a weight sum of quadratic penalties across all structures and objectives, providing a numerical representation of how much each structure's constraints are being violated, with higher penalties reflecting greater deviation from the specified objectives. These two metrics form the foundation for plan evaluation and constraint adjustment. Based on these metrics,  an experienced planner can derive several key insights: (1) the degree of deviation between the current plan and the prescribed clinical goals; (2) whether additional room exists for further plan improvement; and (3) which objectives to adjust and how to do so efficiently to improve plan quality within minimal iterations.

Achieving this level of planning insight requires three core capabilities: (1) arithmetic proficiency to quantify deviations from clinical goals; (2) domain-specific understanding of the optimization system to assess the potential for further improvement; and (3) reasoning ability to interpret trends and current plan status in order to propose targeted constraint adjustments that enhance plan quality effectively and efficiently. Pretrained LLMs possess strong general reasoning capabilities by default but require external support to perform the arithmetic and system-specific evaluation tasks necessary for effective treatment planning \citep{RN34}.
For arithmetic tasks, an  arithmetic tool was developed to compute the numerical deviation between current dosimetric endpoint values, clinical goals, and objective constraint settings. Historical data from all prior iterations, including constraints, dosimetric outcomes, and deviations, were compiled and presented to the LLM, enabling trend-based reasoning and informed decision-making. 

To enable the LLM agent to interpret the optimization environment, relevant domain information about the inverse planning context was encoded into the prompt. This included explanations of key elements such as the meaning and scale of the objective function loss, as well as how constraint deviations relate to potential improvement opportunities. By incorporating this guidance, the agent was better equipped to assess whether a given plan status suggested further room for optimization or, conversely, indicated excessive sparing of certain OARs. Chain-of-thought reasoning \citep{RN35} was employed to enhance the agent’s ability to perform multi-step decision-making, mirroring the logical processes of a human planner. At each iteration, the LLM was prompted to explicitly articulate its reasoning process before proposing new constraint values. This structured reasoning encouraged the model to consider clinical trade-offs, assess constraint violations in context, and prioritize adjustments based on trends observed across prior iterations. By decomposing complex planning decisions into interpretable steps, chain-of-thought prompting improved both the transparency and accuracy of the agent’s constraint refinement.With these supporting components, the LLM agent was able to perform treatment planning in a zero-shot setting, where no prior treatment plans were used to train or fine-tune the agent. 

\subsection{Optimization Setup}

A total of 10 key structures were included in the optimization process: the primary and boost planning target volumes PTV$_{\mathrm{primary}}$ and PTV$_{\mathrm{boost}}$, along with nine OARs: the left and right parotid glands, oral cavity, larynx, pharynx, spinal cord (including a 5~mm margin), brainstem, and mandible.

For the parotids, oral cavity, larynx, and pharynx, patient-specific median dose objectives were defined by the attending radiation oncologist. These values reflect the physician's clinical guidelines, informed by anatomical considerations and patient-specific factors, and serve as guiding references rather than strict constraints. In some cases, no specific numeric objective was assigned to certain structures; in such instances, the LLM agent was expected to exercise independent qualitative judgment regarding how much sparing could be achieved without compromising target coverage.

In contrast, hard constraints were applied to the spinal cord (with margin), brainstem, and mandible, with explicit maximum dose limits intended to ensure protection of critical structures and minimize the risk of severe toxicity. These constraints were expected to be strictly enforced during optimization.

The clinical objective values for each patient were provided to the LLM agent as guidance. However, the agent was not expected to simply replicate these objectives in the optimization engine. Rather, it was tasked with meeting or improving upon the clinical goals when possible, identifying opportunities for enhanced OAR sparing without compromising target coverage or violating hard constraints.

\subsection{Plan Evaluation}

Plan quality was quantitatively assessed by comparing the LLM-generated plans with their corresponding manually generated and clinically approved plans. The evaluated endpoints included the plan-wide maximum dose, the median doses to the bilateral parotid glands, oral cavity, larynx, and pharynx, as well as the maximum doses to critical structures such as the spinal cord (including a 5~mm margin), brainstem, and mandible. In addition to these dose-based metrics, conformity and homogeneity of target coverage were also assessed. 

The Conformity Index (CI) \citep{RN36} was computed as the ratio of the volume covered by the prescription dose ($V_{\mathrm{Pre}}$) to the volume of the planning target ($V_{\mathrm{PTV}}$):  

\begin{equation}
CI = \frac{V_{\mathrm{Pre}}}{V_{\mathrm{PTV}}}.
\end{equation}

The Homogeneity Index (HI) \citep{RN37} was calculated using the formula:  

\begin{equation}
HI = \frac{D_{2} - D_{98}}{D_{\mathrm{pre}}},
\end{equation}

where $D_{2}$ and $D_{98}$ represent the doses received by 2\% and 98\% of the boost PTV, respectively, and $D_{\mathrm{pre}}$ is the prescription dose.To evaluate statistical differences between the LLM-generated and clinical plans, a non-parametric Wilcoxon Signed-Rank test was applied to each dosimetric endpoint, using a significance threshold of $p < 0.05$.
\subsection{Experimental Design}
We utilized two state-of-the-art language models, GPT-4.1 and GPT-4.1-mini, to validate the efficacy of the proposed workflow. GPT-4.1 represents a frontier in OpenAI’s large-scale reasoning model development, offering strong performance across complex multi-step reasoning, mathematical analysis, and domain adaptation tasks. GPT-4.1-mini, while a smaller and more computationally efficient variant, retains core reasoning capabilities but with reduced inference cost and latency. The choice of these two models allows us to benchmark performance across different capacity regimes: GPT-4.1 serving as a high-fidelity upper bound for reasoning quality, and GPT-4.1-mini demonstrating the feasibility of deploying the workflow under more practical computational budgets.

In addition, we designed an ablation study to examine the role of optimization priors in enabling successful application of the workflow. Specifically, the LLM-based agent was tested under two configurations: with optimization priors (including the expected numerical ranges of optimization constraints, tunable parameters, and their directional influence on dose distribution) and without such priors. This design allows us to isolate the contribution of priors in structuring the agent’s reasoning process and guiding constraint refinement. By comparing the agent’s behavior across these conditions, we can assess whether domain-specific grounding is essential for the LLM to successfully conduct automatic inverse planning, as opposed to relying solely on general reasoning ability.

\section{Results}
\label{headings}

\subsection{Dosimetric Endpoints’ Comparison}

The inter-group dosimetric comparison is summarized in Table~\ref{tab:dosimetrics}. Both GPT-4.1 and GPT-4.1-mini with access to optimization priors ( GPT-4.1-WP and GPT-4.1-mini-WP) generated plans of clinically comparable quality, with minimal variation in target coverage and overall OAR sparing relative to the clinical reference. GPT-4.1-WP achieved the most favorable numerical performance across the majority of evaluated metrics, reflecting its stronger reasoning capacity and efficiency in conducting treatment planning. Importantly, the absence of optimization priors resulted in a marked deterioration in planning performance: both GPT-4.1 and GPT-4.1-mini produced plans with significantly worse OAR sparing in the condition without access to optimization priors (WOP), as indicated by elevated OAR doses, compared with their prior-informed counterparts. Although GPT-4.1-mini-WOP yielded slightly lower plan maximum dose and improved conformity indices, these apparent gains stemmed from inadequate OAR protection and thus represent an unfavorable trade-off from a clinical perspective.

\begin{figure}[htbp]
  \centering
  \includegraphics[width=1\linewidth]{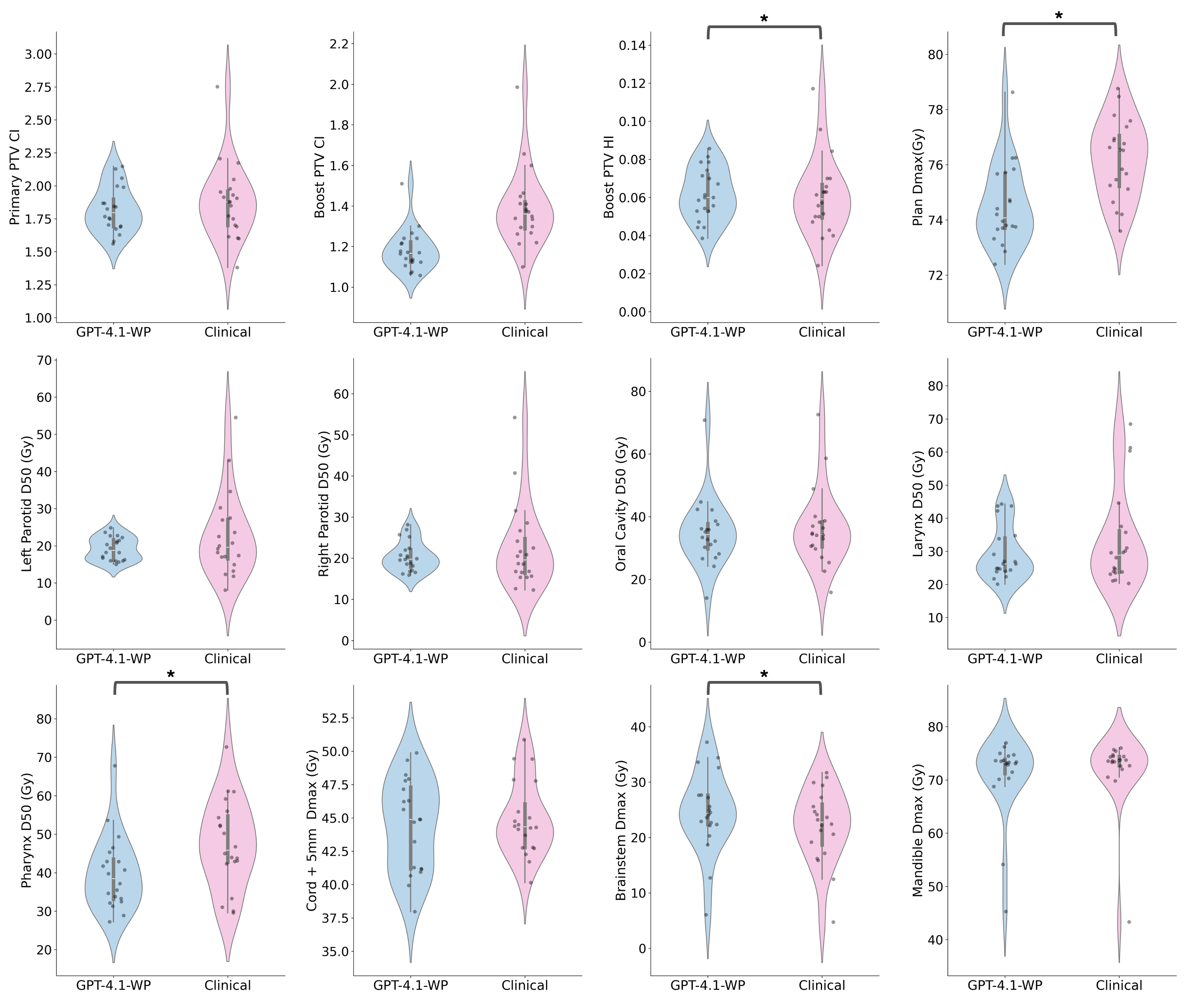}
  \caption{Distribution of dosimetric endpoints for GPT-4.1-WP--generated plans compared with clinical plans. 
  CI: conformity index; HI: homogeneity index; $D_{50}$: median dose. 
  Asterisks ($\ast$) indicate statistically significant differences between groups ($p < 0.05$).}
  \label{fig:violin}
\end{figure}

\begin{table}[htbp]
  \centering
  \caption{Comparison of clinical plans with LLM-generated plans under different configurations. 
  Reported values represent mean ($\pm$ standard deviation) for key dosimetric metrics across all patients. 
  Results are shown for GPT-4.1 and GPT-4.1-mini, with (WP) and without (WOP) access to optimization priors. 
  $D_{\max}$: the maximum dose within the structure; 
  $D_{50}$: the median dose within the structure; 
  CI: conformity index; 
  HI: homogeneity index. 
  For each metric, the optimal value is highlighted in bold.}

  \label{tab:dosimetrics}
  \resizebox{\linewidth}{!}{%
  \begin{tabular}{lccccc}
    \toprule
    & Clinical & GPT-4.1-WP & GPT-4.1-WOP & GPT-4.1-mini-WP & GPT-4.1-mini-WOP \\
    \midrule
    Plan $D_{\max}$ (Gy) & 76.22($\pm$1.44) & 74.53($\pm$1.48) & 74.17($\pm$1.20) & 74.19($\pm$1.07) & \textbf{73.87}($\pm$0.93) \\
    Brainstem $D_{\max}$ (Gy) & \textbf{22.13}($\pm$6.65) & 24.56($\pm$7.21) & 27.57($\pm$7.27) & 24.21($\pm$6.63) & 28.08($\pm$7.26) \\
    Cord + 5mm $D_{\max}$ (Gy) & 44.91($\pm$2.82) & \textbf{44.46}($\pm$3.47) & 48.87($\pm$3.03) & 44.58($\pm$3.97) & 49.59($\pm$3.06) \\
    Mandible $D_{\max}$ (Gy) & 72.06($\pm$6.94) & \textbf{70.86}($\pm$6.94) & 71.66($\pm$6.69) & 71.17($\pm$6.96) & 71.62($\pm$6.42) \\
    Left Parotid $D_{50}$ (Gy) & 22.66($\pm$11.22) & \textbf{19.21}($\pm$3.09) & 23.18($\pm$3.97) & 21.93($\pm$5.71) & 22.99($\pm$3.92) \\
    Right Parotid $D_{50}$ (Gy) & 22.52($\pm$10.17) & \textbf{20.47}($\pm$3.64) & 24.94($\pm$3.75) & 20.70($\pm$5.42) & 25.42($\pm$5.97) \\
    Oral Cavity $D_{50}$ (Gy) & 36.14($\pm$12.44) & 34.95($\pm$10.98) & 38.48($\pm$9.09) & \textbf{33.26}($\pm$11.45) & 39.41($\pm$9.88) \\
    Larynx $D_{50}$ (Gy) & 33.16($\pm$14.42) & \textbf{29.43}($\pm$8.02) & 36.24($\pm$9.36) & 31.29($\pm$9.96) & 37.83($\pm$11.49) \\
    Pharynx $D_{50}$ (Gy) & 47.54($\pm$11.50) & \textbf{39.85}($\pm$9.62) & 49.18($\pm$7.20) & 44.37($\pm$9.04) & 49.43($\pm$8.34) \\
    PTV$_{\mathrm{primary}}$ CI & 1.88($\pm$0.29) & 1.82($\pm$0.17) & 1.92($\pm$0.19) & 1.83($\pm$0.17) & 1.93($\pm$0.17) \\
    PTV$_{\mathrm{boost}}$ CI & 1.39($\pm$0.19) & 1.18($\pm$0.10) & 1.17($\pm$0.09) & 1.17($\pm$0.09) & \textbf{1.16}($\pm$0.09) \\
    PTV$_{\mathrm{boost}}$ HI & 0.061($\pm$0.021) & 0.062($\pm$0.021) & 0.059($\pm$0.020) & 0.058($\pm$0.013) & \textbf{0.055}($\pm$0.019) \\
    \bottomrule
  \end{tabular}
  }
\end{table}
Based on the comparative evaluation, GPT-4.1-WP was selected as the primary model for subsequent analyses, given its consistently superior dosimetric performance and reasoning stability. The distribution of endpoints for clinical plans and GPT-4.1-WP–generated plans are shown in Figure~\ref{fig:violin}. GPT-4.1-WP plans demonstrated lower median values and shorter interquartile ranges for the target conformity index, indicating more consistent target coverage. For parotid sparing, clinical plans showed wider variability, with some cases exhibiting high median doses, whereas GPT-4.1-WP achieved more consistent dose reduction across patients. This discrepancy arises because, in cases with substantial parotid overlap with the target, physicians sometimes omit explicit numerical constraints and instead provide only general guidance to “minimize” dose, leading planners to prioritize the protection of other OARs. In contrast, even without specific parotid constraints, the LLM-agent consistently pursued parotid sparing although moderately and demonstrated reliable improvements across the test cohort. For other OARs, GPT-4.1-WP plans achieved sparing comparable to that of clinical plans, maintaining dose levels within acceptable ranges without compromising target coverage.

\subsection{Case Study and Agent Reasoning}

The clinical constraints for a representative sample case are summarized in Table~\ref{tab:clinical_constraints}. 

\begin{table}[htbp]
  \centering
  \caption{Clinical constraints for a representative example case.}
  \label{tab:clinical_constraints}
  \small
  \resizebox{\linewidth}{!}{%
  \begin{tabular}{ccccccccc}
    \hline
    OAR & Right Parotid & Left Parotid & Oral Cavity & Larynx & Pharynx & Cord+5 mm & Brainstem & Mandible \\
    \hline
    Dose (Gy) & 30--35 & 16 & 35 & 25--30 & Minimize & 45 & 25 & 70 \\
    Volume(absolute or \%)    & Median & Median & Median & Median & -- & Max & Max & Max \\
    \hline
  \end{tabular}%
  }
\end{table}

The constraints are heterogeneous in nature: some are expressed as absolute limits, others as ranges, while certain structures have no explicit numerical constraints. For these latter cases, the directive ``minimize'' indicates that the planner should achieve the lowest dose reasonably achievable. By providing these clinical constraints to the agent, we aim for it to interpret and utilize them in a manner similar to experienced human planners, treating them as reference guidelines and a starting point rather than the ending point, while ultimately striving to achieve optimal plan quality.

\begin{figure}[htbp]
  \centering
  \includegraphics[width=1\linewidth]{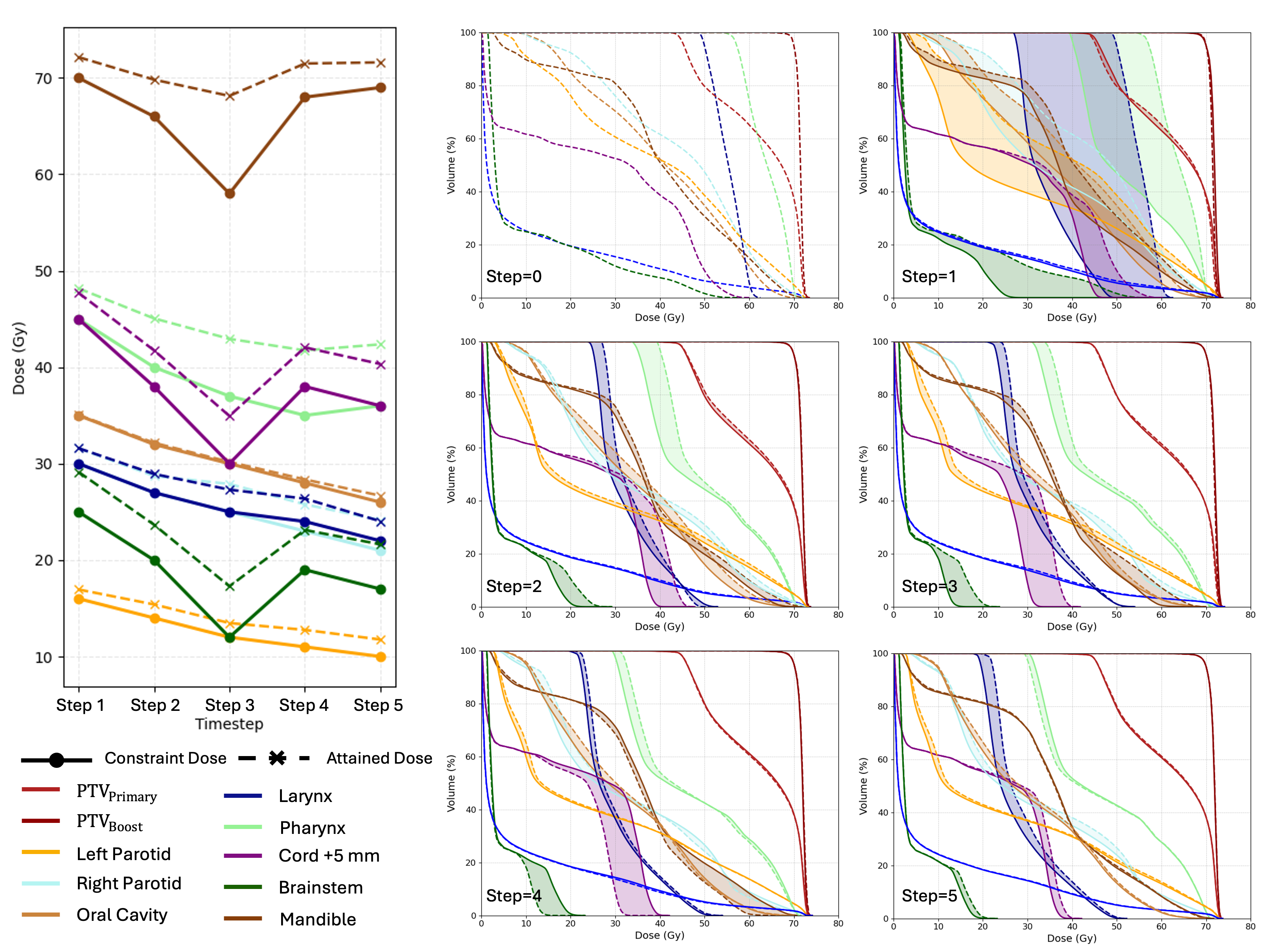}
  \caption{Planning log for the example case. Left panel: trajectories of dose constraints (solid lines) and attained dosimetric endpoints (dashed lines with markers) across optimization steps. Right panels: evolution of DVHs. Step 0 shows the initial plan optimized with PTV constraints only. In subsequent steps, dashed lines indicate the DVHs from the previous step, solid lines represent the updated DVHs, and the shaded regions highlight inter-step DVH changes. }
  \label{fig:track_sum}
\end{figure}

The progression of optimization objective adjustments is illustrated in Figure~\ref{fig:track_sum}. At Step~0, the LLM initialized the optimization by selecting constraint values close to the clinical goals to accelerate convergence. For structures with range-based constraints, the agent chose values at the boundary. For the pharynx, which lacked an explicit numerical constraint, the LLM selected a starting point of 45~Gy median dose, reasoning that this value was ``well below the current dose, but not so low as to risk infeasibility.'' After applying the first set of constraints, the pharynx showed a favorable dose response, prompting the agent to further tighten sparing.
In general, the LLM adopted a strategy of using larger step sizes in the early stages to probe the sparing potential of each structure, followed by smaller step sizes in later stages for fine-tuning and to avoid oversparing. For the mandible, the clinical goal was $D_{\max} < 70$~Gy. The agent attempted to lower the optimization constraint to 58~Gy in pursuit of the clinical objective; however, the attained dose plateaued around 70~Gy and was accompanied by a large increase in the objective function loss. From this history, the LLM reasoned that ``previous steps show that lowering the constraint further increases violation and loss, and actual dose is not decreasing much. This suggests that further sparing is difficult and may compromise PTV coverage.'' Consequently, the agent relaxed the mandible constraint to preserve target coverage. A similar trend was observed for the brainstem and cord+5~mm.

Both the attained dose trajectories and the DVH variations confirm that the LLM agent effectively and efficiently improved plan quality within only a few optimization steps, guided by strong and interpretable reasoning.Planning was performed on a workstation with an Intel Xeon CPU and 32 GB RAM, completing in under 5 minutes, significantly faster than manual planning.

\begin{figure}[htbp]
  \centering
  \includegraphics[width=1\linewidth]{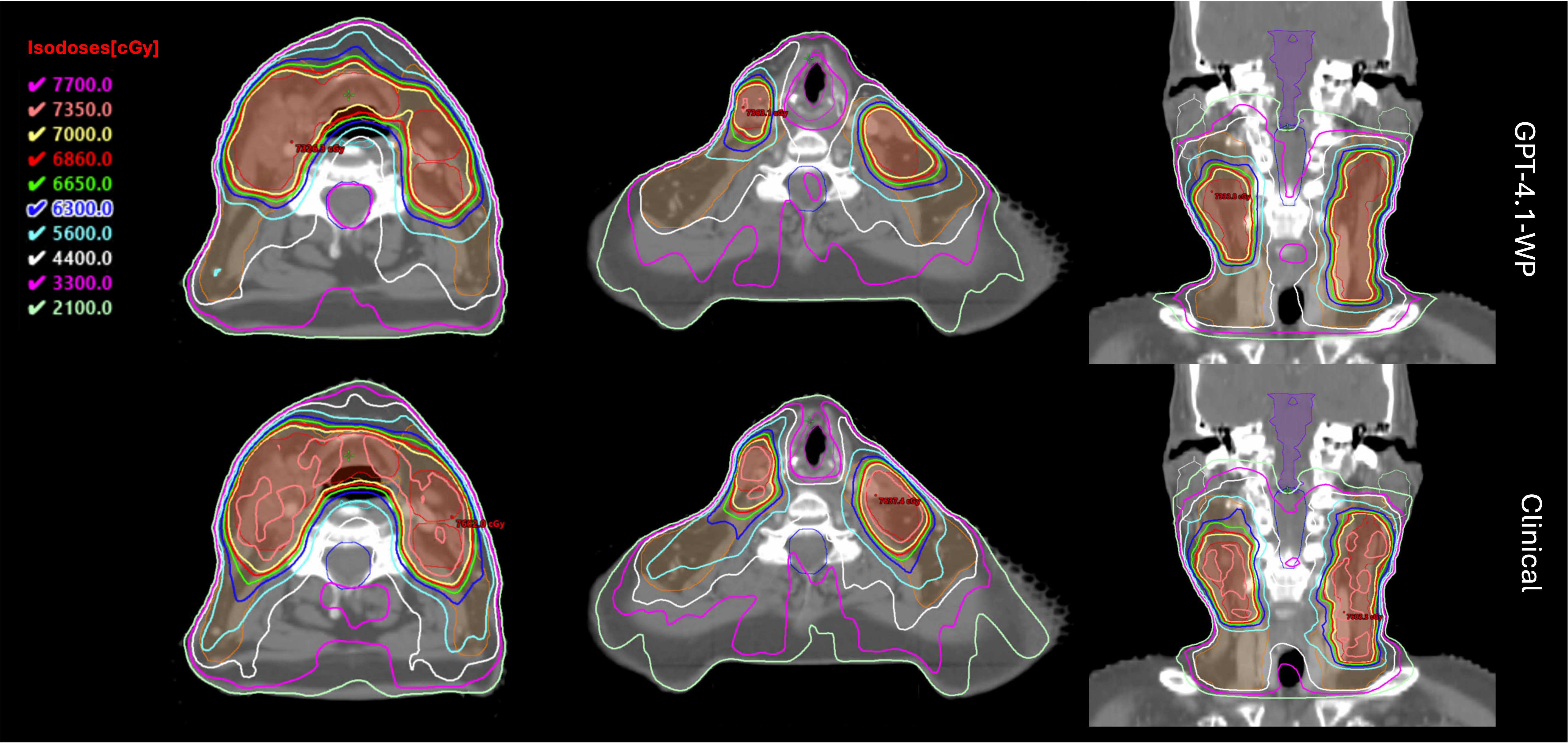}
  \caption{Comparison of isodose distributions between the GPT-4.1-WP–generated plan (left) and the clinical reference plan (right). Red segment: boost PTV; orange segment: primary PTV. }
  \label{fig:isodose}
\end{figure}

The isodose distribution of the LLM-generated plan and the clinical plan is shown in Figure~\ref{fig:isodose}. Both plans achieved adequate coverage of the boost PTV and primary PTV, as evidenced by the 70.0~Gy and 44.0~Gy isodose lines in the axial and sagittal views. In the axial slices of the first column, effective spinal cord sparing is evident in the GPT-4.1-WP plan: the 33~Gy isodose line encircles the cord, highlighting a sharp dose fall-off around this critical structure. A similar pattern of dose fall-off and sparing is observed for the larynx in the axial views of the second column.

In the coronal views (third column), both the LLM-generated and clinical plans demonstrate good conformity of high-dose regions to the target volumes, with the 70.0 Gy and 73.5 Gy isodose lines closely wrapping around the boost PTV. The dose fall-off toward adjacent OARs is well preserved in both plans. In the GPT-4.1-WP plan, the parotid glands exhibit a steeper gradient of intermediate isodose lines, suggesting potential for improved sparing without compromising PTV coverage.
Overall, the alignment of isodose lines between the two plans indicates that the LLM agent was capable of reproducing clinically acceptable dosimetric trade-offs. While target coverage was comparable between the clinical and GPT-4.1-WP plans, the LLM-generated plan demonstrated sharper dose gradients around the spinal cord and larynx.

\section{Discussion}
In this study, the experiments were conducted within a widely adopted commercial treatment planning system (Eclipse\texttrademark, Varian Medical Systems). In contrast to prior work \citep{RN38} that relied on in-house research platforms, embedding the agent directly into a clinical-grade system enhances both generalizability and translational potential. Furthermore, by constraining the agent to the same information available to human planners and restricting its actions to the parameter adjustment space routinely used in practice, we maximized clinical applicability and interpretability.

A notable strength of this work is that the experiments were performed entirely in a zero-shot manner. This is an important step toward generalizable AI-driven planning, particularly for centers where access to large, high-quality training datasets may be limited. Wang et al.\ \citep{RN39} have previously demonstrated a few-shot LLM-based planning approach in lung and cervical cancer, showing feasibility when prior plans are provided. Our work advances this paradigm by eliminating the need for historical plans altogether, thereby reducing bias from training data size and quality.

The results also underscore a critical insight: while LLMs exhibit strong general reasoning capabilities, their clinical utility depends heavily on the information provided. One key component is the interpretation of clinical constraints. In our institution, clinical constraints are determined by attending physicians based on patient assessment and prior experience. These constraints often represent reference values rather than strictly achievable endpoints. For an LLM agent to make clinically relevant decisions, it must be guided to interpret these objectives as flexible reference points rather than absolute targets, mirroring the approach of experienced human planners. Importantly, practices may vary across institutions depending on local clinical guidelines and physician preferences, and thus institution-specific instructions must be provided to ensure that LLM-driven decisions remain clinically relevant.

A clear understanding of the optimization engine is also essential for successful treatment planning.  For example, since the Eclipse engine is driven by a quadratic loss function, effective optimization typically requires setting objectives lower than the desired dose to create a driving force for sparing. The extent of this offset depends on available dosimetric trade-offs and is not intuitive without prior planning experience. Such “hidden rules” are not encoded in the LLM a priori, yet are critical for clinically meaningful outcomes. Clear and structured provision of this knowledge is therefore essential to enable effective autonomous planning, as demonstrated by our results.

By validating this zero-shot workflow in a commercial TPS, we show that LLMs can be deployed across diverse institutions without large training datasets or custom expert systems. This could help alleviate the burden on high-volume centers and provide advanced planning support to smaller centers with limited resources. Nonetheless, this work remains a pilot study. Its efficacy in other disease sites, planning modalities, and clinical environments requires further investigation. Future work will extend this framework to additional tumor sites to validate its robustness and to establish the generalizability of LLM-guided planning beyond head-and-neck cancer.

\section{Conclusion}
In this study, we proposed a clinical ready workflow that leverages an LLM-based agent to perform inverse treatment planning for IMRT within a commercial TPS in a zero-shot setting. Our results demonstrate that the LLM agent can efficiently generate treatment plans with consistent quality, underscoring its potential as a clinically applicable tool for autonomous radiotherapy planning.

{
    \small
    \bibliographystyle{IEEEtran}
    \bibliography{llm_manuscript}

}

%%%%%%%%%%%%%%%%%%%%%%%%%%%%%%%%%%%%%%%%%%%%%%%%%%%%%%%%%%%%

%%%%%%%%%%%%%%%%%%%%%%%%%%%%%%%%%%%%%%%%%%%%%%%%%%%%%%%%%%%%

\newpage

%%% END INSTRUCTIONS %%%

\end{document}